\documentclass[showpacs,twocolumn,aps,preprintnumbers,letterpaper]{revtex4}
\usepackage{amsmath,amssymb}
\usepackage{epsfig}
\usepackage{graphicx}
\usepackage{amsmath}
\usepackage{slashed}
\usepackage{amsfonts}
\usepackage{epstopdf}
\usepackage{color}

\newcommand{\be}{\begin{equation}}
\newcommand{\ee}{\end{equation}}
\newcommand{\bear}{\begin{eqnarray}}
\newcommand{\eear}{\end{eqnarray}}
\newcommand{\ba}{\begin{array}}
\newcommand{\ea}{\end{array}}

\def\be{\begin{eqnarray}}
\def\ee{\end{eqnarray}}
\def\bea{\be}
\def\eea{\ee}

\def\roughly#1{\mathrel{\raise.3ex\hbox{$#1$\kern-.75em%
\lower1ex\hbox{$\sim$}}}}

\definecolor{davecolor}{rgb}{0.95,  0.5,  0.2}

\def\({\left(}
\def\){\right)}
\def\[{\left[}
\def\]{\right]}
\def\<{\langle}
\def\>{\rangle}
\newcommand{\bwt}{\begin{widetext}}
\newcommand{\ewt}{\end{widetext}}

\newcommand{\bi}{\begin{itemize}}
\newcommand{\ei}{\end{itemize}}
\newcommand{\ben}{\begin{enumerate}}
\newcommand{\een}{\end{enumerate}}
\newcommand{\bca}{\begin{cases}}
\newcommand{\eca}{\end{cases}}
\newcommand{\bln}{\begin{align}}
\newcommand{\eln}{\end{align}}
\newcommand{\bst}{\begin{split}}
\newcommand{\est}{\end{split}}

  \long\def\comment#1{ }

  \newcommand{\beq}{\begin{eqnarray}}
  \newcommand{\eeq}{\end{eqnarray}}
  
 \def\simge{\mathrel{%
   \rlap{\raise 0.511ex \hbox{$>$}}{\lower 0.511ex \hbox{$\sim$}}}}
\def\simle{\mathrel{
   \rlap{\raise 0.511ex \hbox{$<$}}{\lower 0.511ex \hbox{$\sim$}}}}

\begin{document}

\title{Strongly coupled $\mathcal{N}=4$ super Yang-Mills plasma on the Coulomb branch I: Thermodynamics}

\author{Kiminad A. Mamo}
\email{kiminad.mamo@stonybrook.edu}
%\email{ismail.zahed@stonybrook.edu}
\affiliation{Department of Physics and Astronomy, Stony Brook University, Stony Brook, New York 11794-3800, USA}
\affiliation{Department of Physics, University of Illinois, Chicago, Illinois 60607, USA}

%\author{Ismail Zahed}
%\email{ismail.zahed@stonybrook.edu}
%\affiliation{Department of Physics and Astronomy, Stony Brook University, Stony Brook, New York 11794--3800, USA}

%%%%%%%%%

\date{\today}
\begin{abstract}
We study $\mathcal{N} = 4$ super Yang-Mills theory on the Coulomb branch (cSYM) in the strong coupling limit by using the AdS/CFT correspondence. The dual geometry is the rotating black 3-brane Type IIB supergravity solution with a single non-zero rotation parameter $r_{0}$ which sets a fixed mass scale corresponding to the scalar condensate $<\mathcal{O}>\,\,\sim r_{0}^4$ in the Coulomb branch. We introduce a new ensemble where $T$ and $<\mathcal{O}>$ are held fixed, i.e., the free energy $F(T,<\mathcal{O}>)$ is a function of $T$ and $<\mathcal{O}>$. We compute the equation of state (EoS) of $\mathcal{N} = 4$ cSYM at finite $T$, as well as the heavy quark-antiquark potential and the quantized mass spectrums of the scalar and spin-2 glueballs at $T=0$. By computing the Wilson loop (minimal surface) at $T=0$, we determine the heavy quark-antiquark potential $V(L)$ to be the Cornell potential, which is confining for large separation $L$. At $T\neq 0$, we find two black hole branches: the large black hole and small black hole branches. For the large black hole branch, that has positive specific heat, we find qualitatively similar EoS to that of pure Yang-Mills theory on the lattice. For the small black hole branch, that has negative specific heat, we find an EoS where the entropy and energy densities decrease with $T$. We also find a second-order phase transition between the large and small black hole branches with critical temperature $T_c=T_{min}$.
\end{abstract}
%\pacs{11.25.Tq, 11.15.Kc, 12.38.Lg}
%\pacs{11.25.Tq, 13.60.Hb,13.85.Lg}
%11.25.Tq 	Gauge/string duality  13.60.Hb for deep-inelastic structure functions; 13.85.Lg 	Total cross sections

%11.15.Kc	Classical and semiclassical techniques
%11.30.Rd	Chiral symmetries
%12.38.Lg	Other nonperturbative calculations

\maketitle

\setcounter{footnote}{0}

%\baselineskip 18pt \pagebreak
%\renewcommand{\thepage}{\arabic{page}}
%\tableofcontents
%\pagebreak

\section{Introduction}
The AdS/CFT correspondence \cite{Maldacena:1997re, Gubser:1998bc, Witten:1998qj} has opened a new window to the strongly coupled regime of gauge theories such as $\mathcal{N}=4$ super Yang-Mills (SYM). Unfortunately, so far, we luck an exact string theory dual to QCD even though there are various works which explored different non-conformal deformations of $\mathcal{N}=4$ SYM both on the top-down (where both the details of the deformation of $\mathcal{N}=4$ SYM and its string theory dual are known) \cite{Erlich:2005qh, deTeramond:2005su, Karch:2006pv, Liu:2008tz, Chelabi:2015gpc, Gubser:2008yx, Gubser:2008sz, Rougemont:2015wca, Gursoy:2010fj}, and bottom-up approaches (where the details of the deformation of $\mathcal{N}=4$ SYM and its string theory dual are unknown) \cite{Karch:2002sh, Mateos:2006nu, Erdmenger:2007cm, Witten:1998zw, Sakai:2004cn, Rebhan:2014rxa, Witten:1998zw, Kraus:1998hv, Brandhuber:1999jr, Cvetic:1999xx, Bakas:1999ax, Klebanov:2000hb, Polchinski:2000uf, Maldacena:2000yy}.

%In this paper, we extend $\mathcal{N}=4$ cSYM at zero temperature to non-zero temperature.
In $\mathcal{N}=4$ SYM on the Coulomb branch (cSYM) at zero temperature, a scale is introduced dynamically through the Higgs mechanism where the scalar particles $\Phi_{i}$ (i=1...6) of $\mathcal{N}=4$ SYM acquire a non-zero vacuum expectation value (VEV) that breaks the conformal symmetry, and the gauge symmetry $SU(N_{c})$ to its subgroup $U(1)^{N_{c}-1}$ without breaking the supersymmetry, and without resulting in a running of the coupling constant \cite{Kraus:1998hv}. At finite temperature, the mechanism is the same except the fact that supersymmetry will be broken as well.

The string theory dual for $\mathcal{N}=4$ cSYM at zero temperature is well known. Among various Type IIB supergravity background solutions that are dual to the strongly coupled $\mathcal{N}=4$ cSYM at zero temperature \cite{Kraus:1998hv, Brandhuber:1999jr, Cvetic:1999xx, Bakas:1999ax}, in this Letter, we will study a Type IIB supergravity background solution that describes non-extremal rotating black 3-branes (with mass parameter $m$ and single rotational parameter $r_{0}$) which, in the extremal limit, i.e., $r_{0}\gg m^{1/4}$, is dual to $\mathcal{N}=4$ SYM on the Coulomb branch at zero temperature that arises from $N_{c}$ D3-branes distributed uniformly in the angular direction, inside a 3-sphere of radius $r_{0}$ \cite{Brandhuber:1999jr}.

So far the studies of the non-extremal rotating black 3-brane supergravity backgrounds has been limited to the grand canonical ensemble (which is described by fixed temperature $T$ and angular velocity $\Omega$ or chemical potential $\mu$, i.e. the Gibbs free energy $G(T,\mu)$ is a function of $T$ and $\mu$), and canonical ensemble (which is described by fixed temperature $T$ and angular momentum density $J$ or charge density $<J^{0}>\,\,=\rho$, i.e., the Helmholtz free energy $F(T,<J^{0}>)$ is a function of $T$ and $<J^{0}>$), see \cite{Gubser:1998jb, Behrndt:1998jd, Cvetic:1999ne, Cvetic:1999rb, Cai:1998ji, Chamblin:1999tk, Avramis:2006ip, Son:2006em, DeWolfe:2011ts, Wu:2014xva, Finazzo:2016psx}. The two ensembles have different physics, for example, in planar rotating black 3-branes, Hawking-Page phase transition does not exist in the grand canonical ensemble even though it does exist in the canonical ensemble \cite{Cai:1998ji, Wu:2014xva}.

In this paper, we will introduce a new ensemble which is described by a fixed temperature $T$ and a scalar condensate $<\mathcal{O}>$, i.e., the Helmholtz free energy $F(T,<\mathcal{O}>)$ is a function of $T$ and $<\mathcal{O}>$. The scalar condensate is the expectation value of dimension 4 operator $\mathcal{O}=Tr\Phi_{i_{1}}\Phi_{i_{2}}\Phi_{i_{3}}\Phi_{i_{4}}$, that is, $<\mathcal{O}>\sim \lim_{r\rightarrow\infty}\sqrt{-g}g^{rr}\partial_{r}h\sim\Lambda^4$ of the massless metric fluctuation $h=\bar{g}^{\mu\nu}h_{\mu\nu}=1-\bar{g}^{\mu\nu}g_{\mu\nu}$, where $\bar{g}_{\mu\nu}$ is the metric component of pure $AdS_{5}\times S^5$ space while $g_{\mu\nu}$ is our 10-dimensional metric (\ref{10metric}) \cite{Kraus:1998hv}, and $\Lambda\equiv\frac{r_{0}}{\pi R^2}$ with $R$ the radius of the $AdS_{5}$ space. 

Therefore, in our ensemble, the variation of the Helmholtz free energy $F(T,<\mathcal{O}>)$ can be written as
\begin{equation}\label{FO}
dF(T,<\mathcal{O}>)= -SdT+h_{0}d<\mathcal{O}>
\end{equation}
where the source $h_{(0)}=h(r\rightarrow\infty)$. One can compare the variation of the free energy in our ensemble (\ref{FO}) to the variation in the canonical ensemble \begin{equation}
dF(T,<J^{0}>)= -SdT+A_{t}^{(0)}d<J^{0}>
\end{equation}
where the source $A_{t}^{(0)}=A_{t}(r\rightarrow\infty)=\mu$, and grand canonical ensemble
\begin{equation}
dG(T,\mu)= -SdT-<J^{0}>d\mu\,.
\end{equation}

The outline of this paper is as follows: In section \ref{thermo}, we study the thermodynamics of rotating black 3-brane solution where a single rotation parameter $r_{0}$ is turned on. In section \ref{Cornell}, we compute the heavy quark-antiquark potential $V(L)$ of $\mathcal{N}=4$ cSYM. In section \ref{glueballs}, we study the mass spectrum of glueballs in $\mathcal{N}=4$ cSYM. 

\section{\label{thermo}Thermodynamics of $\mathcal{N}=4$ cSYM plasma}
The rotating black 3-brane solution of the 5-dimensional Einstein-Maxwell-scalar action found from the $U(1)^3$ consistent truncation of Type IIB supergravity on $S^5$ \cite{Cvetic:1999xp, Cvetic:2000nc}, see also \cite{Donos:2011qt, Mamo:2016oli, Mamo:2015aia}, is given by
\begin{equation}
ds_{(5)}^2 = \frac{r^2}{R^2}{H}^{1/3}\Big(-f\, dt^2+dx^2 + dy^2 + dz^2\Big)
+\frac{ H^{-2/3}}{\frac{r^2}{R^2}f}dr^2\,,
\label{metric}
\end{equation}
where
\begin{equation}
f= 1-\frac{r^4_h}{r^4}\frac{H(r_{h})}{H(r)}\,,\,\,\,H=1-\frac{r_{0}^2}{r^2}\,,
\end{equation}
\bea
\varphi_{1} &=& \frac{1}{\sqrt{6}}\ln H\,,\,\, \varphi_{2} = \frac{1}{\sqrt{2}}\ln H  \,,\nonumber\\
A^1_t &=&i\frac{r_{0}}{R^2}\frac{r_{h}^2\sqrt{H(r_{h})}}{r^2H(r)}\,,\nonumber\\
r_{h}^2 &=&\frac{1}{2}\Big(r_{0}^2+\sqrt{r_{0}^4+4m}\Big)\,,
\eea
$\kappa=\frac{r_{0}^2}{r_{h}^2}$, $m$ is the mass parameter, and $A^2_t=A^3_t=0$. Note that our metric (\ref{metric}) is equivalent to the metric used in \cite{Son:2006em} after analytically continuing $r_{0}\rightarrow -i\sqrt{q}$. We should also note that having an imaginary gauge potential, in our ensemble, doesn't lead to any inconsistencies, since all physical quantities in the 5-dimensional spacetime are given in terms of $(\partial_{r}A^1_{t})^2$. From the field theory side, having an imaginary gauge potential or imaginary chemical potential $\mu$, means that we are studying the phase diagram of $\mathcal{N}=4$ cSYM at finite $T$ and imaginary chemical potential which is similar to studying the phase diagram of QCD at finite $T$ and imaginary chemical potential which is well known that it doesn't lead to any inconsistencies, see \cite{Roberge,Alford:1998sd,DElia:2002tig,deForcrand:2002hgr} for the study of lattice QCD at finite imaginary chemical potential. 

\begin{figure}
 \begin{center}
\includegraphics[width=0.48\textwidth]{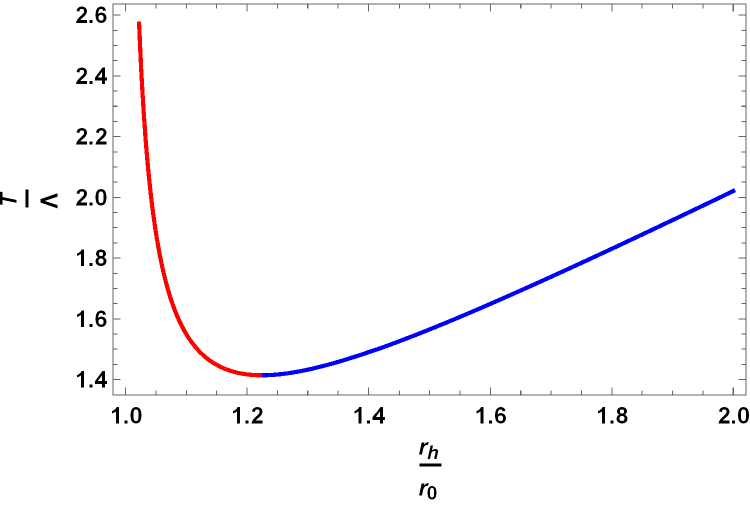}
\caption{Hawking temperature (\ref{T}).}\label{fT}
 \end{center}
\end{figure}

The Hawking temperature $T$ of the black hole (rotating black 3-brane) solution (\ref{metric}) is given by
\begin{equation}\label{T}
\frac{T}{\Lambda} = {1-\frac{1}{2}\kappa\over
\sqrt{\kappa-\kappa^2}}\,\,,
\end{equation}
where $T_{0}=\frac{r_{h}}{\pi R^2}$, $\Lambda=\frac{r_{0}}{\pi R^2}$, and $\kappa=\frac{r_{0}^2}{r_{h}^2}=\frac{\Lambda^2}{T_{0}^2}$. We have plotted $\frac{T}{\Lambda}$ in Fig.~\ref{fT}. We can also invert (\ref{T}) to find
\begin{equation}\label{k}
 \kappa=\frac{1+\frac{T^2}{\Lambda^2}\Big(1\mp\sqrt{\frac{T^2}{\Lambda^2}-2}\Big)}{\frac{1}{2}+2\frac{T^2}{\Lambda^2}}\,.
\end{equation}
Note that in (\ref{k}) $"-"$ corresponds to large black hole branch and $"+"$ corresponds to small black hole branch.

\begin{figure}
 \begin{center}
\includegraphics[width=0.48\textwidth]{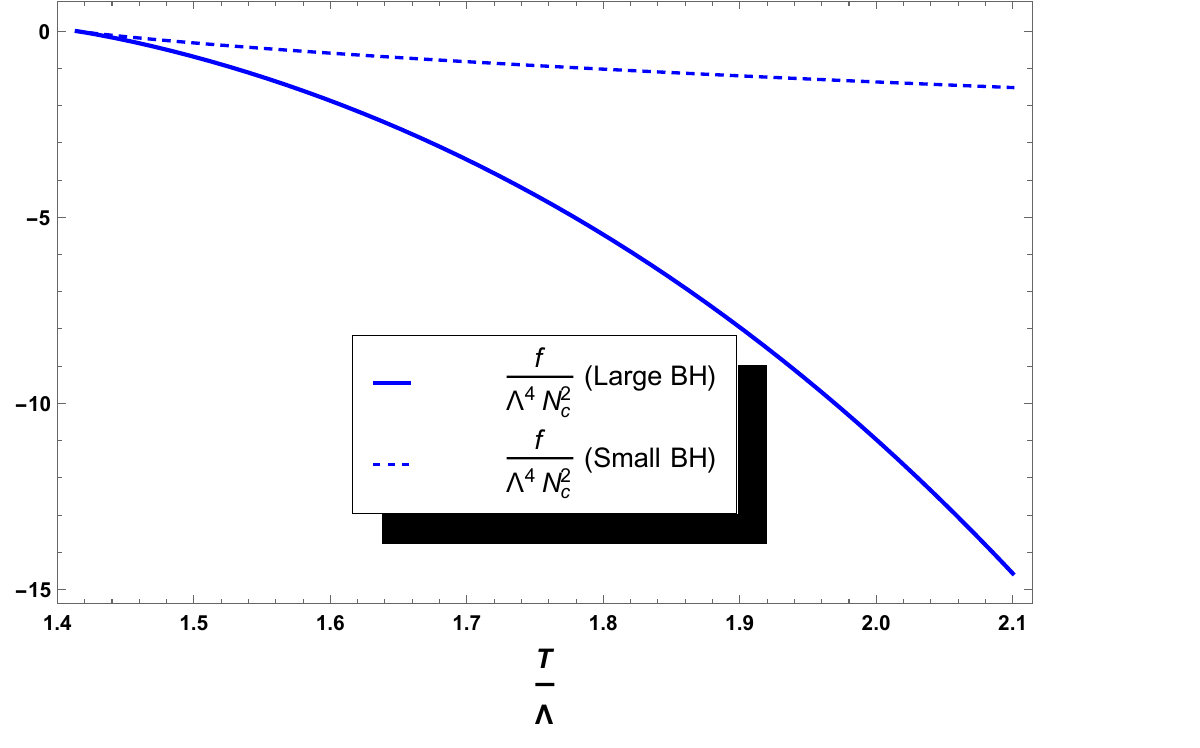}
\caption{The free energy density $\frac{f}{\Lambda^4 N_{c}^2}$ of $\mathcal{N}=4$ cSYM plasma (\ref{f2}) for the large and small black holes.}\label{ff}
 \end{center}
\end{figure}

The entropy density $s(T,\Lambda)$, for our ensemble where $T$ and $\Lambda$ are held fixed, is given by
\begin{eqnarray}
s(T,\Lambda)&=& {A_H\over 4 G_5 V_3}=\frac{1}{4G_{5}}\sqrt{g_{xx}(r_{h})g_{yy}(r_{h})g_{zz}(r_{h})}\,\nonumber\\
&=& {\pi^2 N_c^2 T_{0}^3 \over 2}(1-\kappa)^{1/2}\,,
\label{entropy_density}
\end{eqnarray}
where $G_5=\pi R^3/2 N_c^2$, and $V_{3}$ is the three-dimensional volume. And, using  (\ref{FO}) for fixed $<\mathcal{O}>\sim\Lambda^4$, the corresponding free energy density $f(T,\Lambda)$ of our ensemble can be determined by integrating the entropy density $s(T,\Lambda)$ as \cite{Gubser:2008yx, Gursoy:2010fj}
\begin{eqnarray}\label{f2}
  f(T,\Lambda)&=&-\int^{r_{h}}_{r_{hmin}}\frac{dT'}{dr'_{h}}s(r'_{h},\Lambda)dr'_{h}\,\nonumber\\
              &=&-{\pi^2 N_c^2 T_{0}^4 \over 8}(1-\kappa-\frac{3}{4}\kappa^2-\kappa^2\log(\frac{2}{\kappa}-2))\,,\nonumber\\
\end{eqnarray}
where we choose $r_{hmin}=\sqrt{\frac{3}2{}}r_{0}$, and set the integration constant $f(T_{min},\Lambda)=0$. We have plotted the free energy density $f(T,\Lambda)$ (\ref{f2}) in Fig.~\ref{ff}.

The other thermodynamic quantities can be determined from the free energy density $f(T,\Lambda)$ (\ref{f2}) as: pressure $p=-f$, energy density $\epsilon=p+Ts$, specific heat $C_{\Lambda}=T\Big(\frac{\partial s}{\partial T}\Big)_{\Lambda}$, and speed of sound $c_{s}^2=\frac{\partial p}{\partial\epsilon}=\frac{s}{C_{\Lambda}}$. We have plotted the thermodynamics quantities in Fig.~\ref{fEoS}, Fig.~\ref{facsym}, Fig.~\ref{Ccsym}, Fig.~\ref{cscsym}. To compare our results with pure Yang-Mills theory on the lattice and improved holographic QCD see Fig.5-9 in \cite{Gursoy:2010fj}.

\begin{figure}
 \begin{center}
\includegraphics[width=0.48\textwidth]{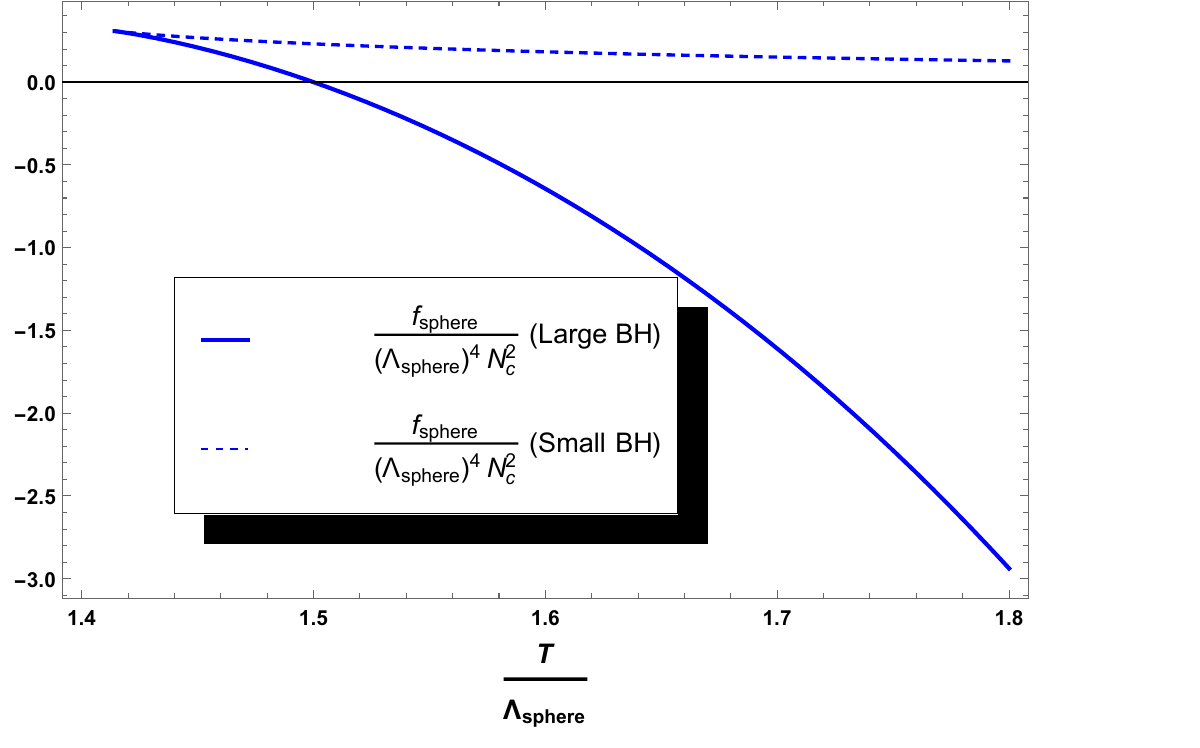}
\caption{The free energy density $\frac{f_{sphere}}{(\Lambda_{sphere})^4 N_{c}^2}$ of $\mathcal{N}=4$ SYM plasma on 3-sphere of radius $R$ (\ref{fsphere}) for the large and small black holes.}\label{ffs}
 \end{center}
\end{figure}

\begin{figure}
 \begin{center}
\includegraphics[width=0.48\textwidth]{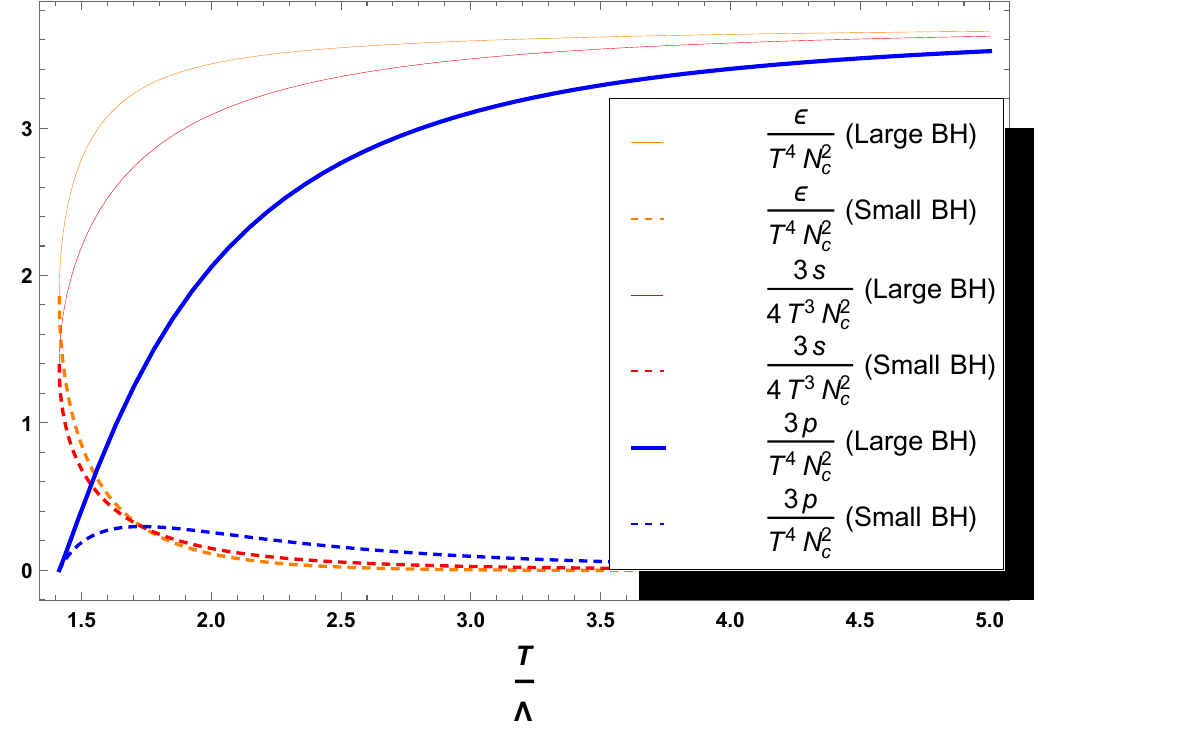}
\caption{The energy density $\frac{\epsilon}{T^4}$, entropy density $\frac{3}{4}\frac{s}{T^3N_{c}^2}$, and pressure $\frac{3p}{T^4N_{c}^2}$ of $\mathcal{N}=4$ cSYM plasma for the large and small black holes.}\label{fEoS}
 \end{center}
\end{figure}

\begin{figure}
 \begin{center}
\includegraphics[width=0.48\textwidth]{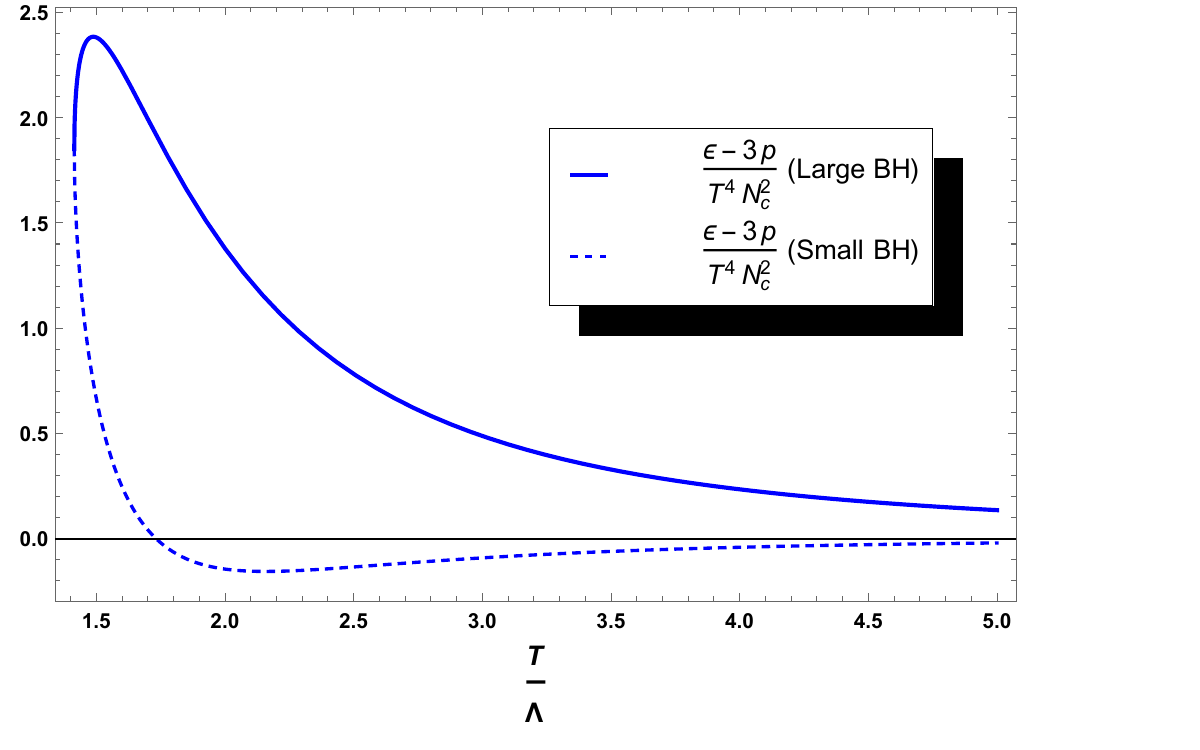}
\caption{The trace anomaly $\frac{\epsilon-3p}{T^4N_{c}^2}$ of $\mathcal{N}=4$ cSYM plasma for the large and small black holes.}\label{facsym}
 \end{center}
\end{figure}

\begin{figure}
 \begin{center}
\includegraphics[width=0.48\textwidth]{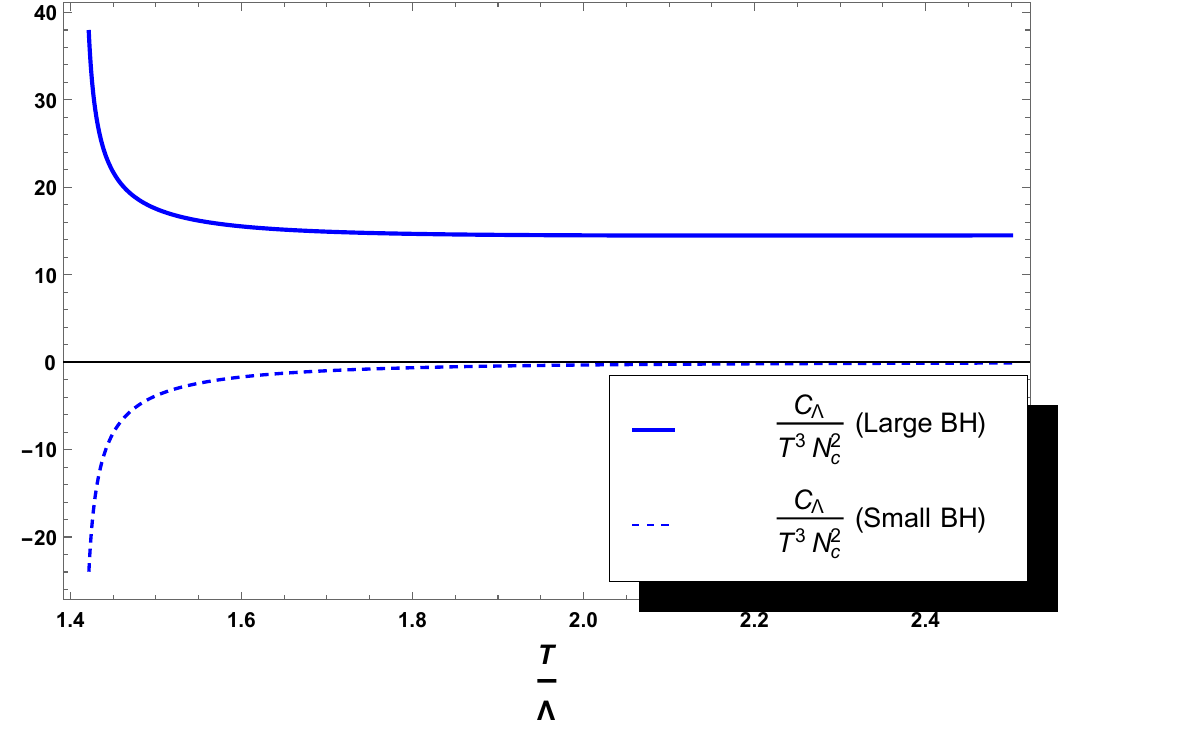}
\caption{The specific heat $C_\Lambda$ of $\mathcal{N}=4$ cSYM plasma for the large and small black holes.}\label{Ccsym}
 \end{center}
\end{figure}

\begin{figure}
 \begin{center}
\includegraphics[width=0.48\textwidth]{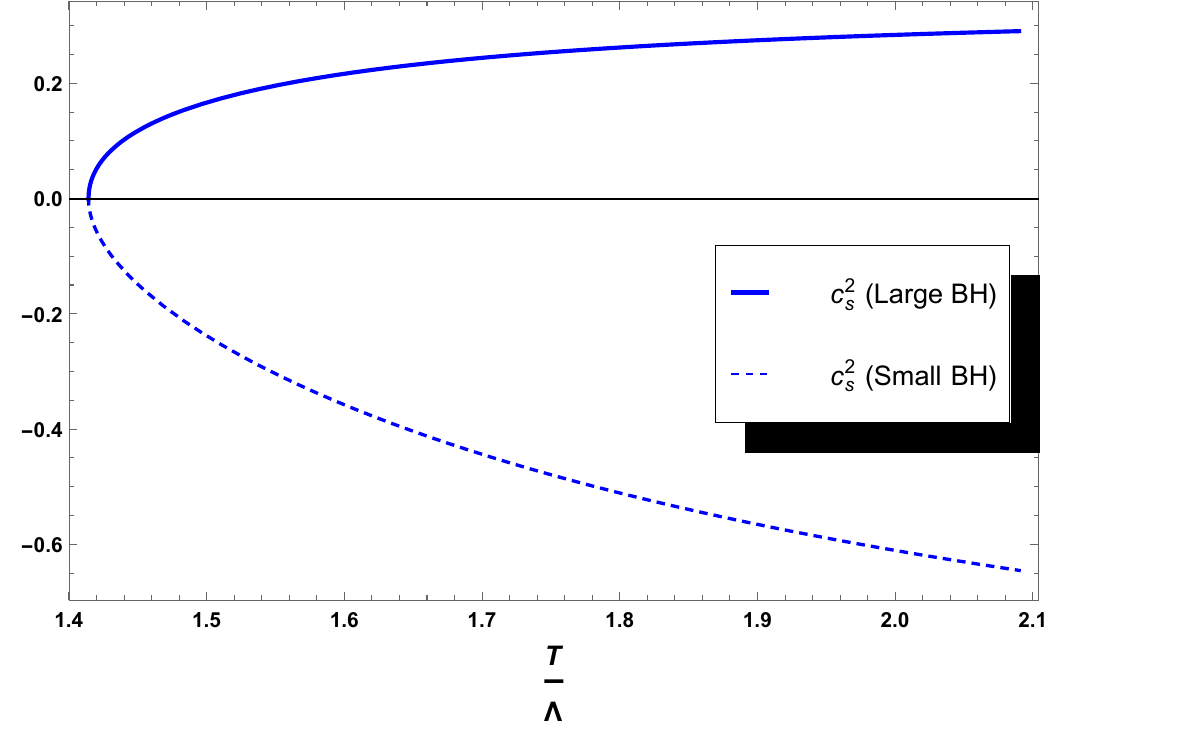}
\caption{The speed of sound $c_s^2$ of $\mathcal{N}=4$ cSYM plasma for the large and small black holes.}\label{cscsym}
 \end{center}
\end{figure}

As a comparison to $\mathcal{N}=4$ cSYM, we have also plotted, see Fig.~\ref{ffs}, the free energy density of $\mathcal{N}=4$ SYM on sphere $f_{sphere}$, which is given by \cite{Witten:1998zw}, see also \cite{Natsuume:2014sfa},
\begin{equation}\label{fsphere}
  f_{sphere}=\frac{F_{sphere}}{V_{3}}=-{\pi^2 N_c^2 T_{0}^4 \over 8}(1-\kappa_{sphere})\,,
\end{equation}
where $\kappa_{sphere}=\frac{R^2}{r_{h}^2}=\frac{\Lambda_{sphere}^2}{T_{0}^2}$ with $\Lambda_{sphere}=\frac{1}{\pi R}$ and $T_{0}=\frac{r_{h}}{\pi R^2}$, and the Hawking temperature $\frac{T}{\Lambda_{sphere}}=\frac{1+\frac{1}{2}\kappa_{sphere}}{\sqrt{\kappa_{sphere}}}$.

Comparing Fig.~\ref{ff} and Fig.~\ref{ffs}, one can see that we have a second-order phase transition for $\mathcal{N}=4$ cSYM at $T_c=T_{min}$ (it is second-order since the second-derivative of our order parameter (the free energy) or its specific heat capacity is discontinuous, see Fig.~\ref{Ccsym}, while the first derivative of its free energy or the entropy is continuous as one goes from the large black hole to small black hole, see Fig.~\ref{fEoS}, also see \cite{DeWolfe:2011ts, Finazzo:2016psx} for similar second-order phase transitions between large and small black holes in $\mathcal{N}=4$ SYM at finite-chemical potential). And we have a first-order (Hawking-Page) phase transition in $\mathcal{N}=4$ SYM on sphere at $T_c=\Lambda_{sphere}=\frac{1}{\pi R}$ (it is first-order since the first derivative of the free energy or its entropy changes dicontinuously as one goes from the large black hole with $s\sim N_c^2$ to the thermal-AdS with $s\sim 0$).

\section{\label{Cornell}Cornell potential in $\mathcal{N}=4$ cSYM}
The Nambu-Goto (NG) action is
\be \label{NGpar}
S_{NG}=\int d\tau d\sigma \mathcal{L}(h_{ab})=-\frac{1}{2\pi\alpha'}\int d\tau d\sigma \sqrt{-det\,h_{ab}}\,,\nonumber\\
\ee
where the background induced metric on the string $h_{ab}=g_{\mu\nu}\partial_{a}x^{\mu}(\tau,\sigma)\partial_{b}x^{\nu}(\tau,\sigma)$. Using the embedding $(\tau, \sigma)\Rightarrow (t(\tau, \sigma),0,0,x(\tau, \sigma), r=\sigma)$, the background induced metric $h_{ab}(x')$ becomes (${'} \equiv d/d\sigma$)
\begin{eqnarray} \label{bgindz4}
h_{\tau\tau}(x')&=&g_{tt}\,,\nonumber\\
h_{\sigma\sigma}(x')&=&g_{rr}\Bigg(\frac{1}{1+\frac{C^2}{g_{xx}g_{tt}}}\Bigg)\,,
\end{eqnarray}
where we used
\be\label{sol2}
(x')^2=\frac{-C^2g_{rr}}{g^2_{xx}g_{tt}}\frac{1}{\big(1+\frac{C^2}{g_{tt}g_{xx}}\big)}\,.
\ee
which is the solution of the NG equation of motion, and the integration constant $C$ is related to the conjugate momenta $\Pi=\frac{\partial\mathcal{L}}{\partial x'}=-\frac{C}{2\pi\alpha'}$.

Considering a string configuration where a heavy quark is attached to each ends of the string, we can extract the potential energy $V(L)$, of the two quarks separated by length $L$, from the on-shell Nambu-Goto action $S_{NG}$ as
\begin{equation}\label{VL}
  V(L)=\frac{-2S_{NG}}{\mathcal{T}}\,,
\end{equation}
where
\begin{eqnarray}\label{}
  -\frac{2\pi\alpha'}{\mathcal{T}}S_{NG}&=&\int_{r_{m}}^{\infty} dr \Big(\sqrt{-det\,h_{ab}(x')}-\sqrt{-det\,h_{ab}(0)}\Big)\,\nonumber\\
  &-&\int_{r_{h}}^{r_{m}} dr \sqrt{-det\,h_{ab}(0)}\,,
\end{eqnarray}
and $r_{m}$ is related to $L$ through the boundary condition $\frac{L}{2}=\int_{r_{m}}^{\infty}x'\,dr$, and we also fix the integration constant $C$ by demanding $x'\mid_{r=r_{m}}\rightarrow \infty$ which is satisfied only when $C^2=-g_{tt}(r_{m})g_{xx}(r_{m})$. Note that we have a factor of 2 in (\ref{VL}) because our gauge covers only half of the full string configuration which accounts to only half of the full potential energy between the quarks, see \cite{Natsuume:2014sfa} for discussion on how to compute $V(L)$ in the $x(r)$ gauge instead of the widely used $r(x)$ gauge of \cite{Maldacena:1998im}.

For $r\gg r_{m}$, after approximating $h_{\sigma\sigma}(x')\cong h_{\sigma\sigma}(0)= g_{rr}$,
\begin{eqnarray}\label{vl}
  V(L)&\simeq&-\frac{1}{\pi\alpha'}\int_{r_{0}}^{r_{m}} dr \sqrt{-det\,h_{ab}(0)}\nonumber\\
&\simeq&-\frac{2\sqrt{\lambda}}{3\pi}\frac{1}{L}+\frac{\pi\sqrt{\lambda}\Lambda^2}{4}L+\frac{5\sqrt{\lambda}\Lambda}{6}+\mathcal{O}(r_{0}^4)\,,\nonumber\\
\end{eqnarray}
where we used $\frac{L}{2}=\int_{r_{m}}^{\infty}x'\,dr\cong\frac{1}{3}\frac{R^2}{r_{m}}$ with $x'\cong\frac{g_{xx}(r_{m})}{g_{xx}}\sqrt{\frac{g_{rr}}{g_{xx}}}\cong\frac{r_{m}^2R^2}{r^4}$ for $r\gg r_{m}$, and we have set $r_{h}=r_{0}$ and $f=1$ in the extremal limit.

In \cite{Brandhuber:1999jr}, the heavy quark-antiquark potential energy $V(L)$ was computed for the 10-dimensional background metric (\ref{10metric}) after analytically continuing $t\rightarrow -it$ and in the extremal limit where $r_{h}=r_{0}$ or $\tilde{f}=f=1$ case. The authors have shown that, for $\theta=\frac{\pi}{2}$, $V(L)$ smoothly interpolates between a Coulombic potential $V(L)=-\frac{2\Gamma(3/4)^2\sqrt{\lambda}}{\Gamma(1/4)^2}\frac{1}{L}$ for small L and a confining potential $V(L)=\frac{\pi\sqrt{\lambda}\Lambda^2}{2}L$ for large $L$. See curve (b) in Fig.5 of \cite{Brandhuber:1999jr}. Their numerical result agrees qualitatively with our analytic result (\ref{vl}) on the 5-dimensional metric (\ref{metric}).

\section{\label{glueballs}Glueballs in $\mathcal{N}=4$ cSYM}
It can easily be shown that bulk fluctuations in the 5-dimensional metric (\ref{metric}), at least in the near boundary limit where the metric is essentially $AdS_{5}$ space with IR cut-off at $r=r_{0}$, have mass-gap and quantized mass spectrum proportional to $\Lambda=\frac{r_{0}}{\pi R^2}$.

In \cite{Brandhuber:1999jr}, it was shown that a scalar bulk fluctuation in a 10-dimensional metric (which is the 10-dimensional uplift of (\ref{metric}))
\begin{eqnarray}
ds_{(10)}^2 &=& \frac{r^2}{R^2}{\tilde{H}}^{1/2}\Big(-\tilde{f}\, dt^2+dx^2 + dy^2 + dz^2\Big)\,\nonumber\\
&+&\frac{\tilde{H}^{1/2}H^{-1}}{\frac{r^2}{R^2}f}dr^2+R^2\Big(\tilde{H}^{1/2}d\theta^2+H\tilde{H}^{-1/2}\sin^2\theta d\phi^2\,\nonumber\\
&+&\tilde{H}^{-1/2}\cos^2\theta d\Omega_{3}^2\Big)+2A_{t}^{1}H\tilde{H}^{-1/2}R^2\sin^2\theta dtd\phi\,,\nonumber\\
\label{10metric}
\end{eqnarray}
where
\begin{equation}\label{}
\tilde{H}=\sin^2\theta+H\cos^2\theta\,,\,\,\text{and}\,\,\,\tilde{f}=1-\frac{r_{h}^4}{r^{4}}\frac{H(r_{h})}{\tilde{H}(r)}\,,
\end{equation}
$f$ and $H$ are the same as in (\ref{metric}), after analytically continuing $t\rightarrow -it$, indeed has mass gap proportional to $\Lambda$ and a quantized mass spectrum $M_{n}^2=4\pi^2\Lambda^2 n(n+1)$, see Eq.54 in \cite{Brandhuber:1999jr}. Since, a scalar bulk fluctuation in (\ref{10metric}) has the same 5-dimensional bulk equation of motion as in \cite{Brandhuber:1999jr} which is the Jacobi equation, we can use this result to calculate the mass spectrum of glueballs in $\mathcal{N}=4$ cSYM.

The transverse gravitational tensor fluctuation $h^{x}_{y}(t,z,r)$ in the 10-dimensional metric (\ref{10metric}), which is a source to dimension 4 stress-energy tensor operator $T^{y}_{x}$, also has the same 5-dimensional bulk equation of motion as the scalar field which is the Jacobi equation. Therefore, we can infer that the operator $T^{y}_{x}$ which corresponds to spin-2 glueballs of $J^{PC}=2^{++}$ \cite{Constable:1999gb} has mass spectrum given by $M_{n}^2=4\pi^2\Lambda^2 n(n+1)$ for $n=1,2,...$.

The real and imaginary parts of the bulk fluctuation of a massless complex scalar field $\Phi=e^{-\phi}+i\chi$, in the 10-dimensional metric (\ref{10metric}), are sources to the dimension 4 scalar operators $\mathcal{O}_{4}=Tr\,F^2$ and $\mathcal{\tilde{O}}_{4}=Tr\,F\wedge F$, respectively \cite{Csaki:1998qr}, and its 5-dimensional bulk equation of motion is the Jacobi equation. Therefore, $\mathcal{O}_{4}$ and $\mathcal{\tilde{O}}_{4}$ which correspond to the scalar glueballs of $J^{PC}=0^{++}$ and $J^{PC}=0^{-+}$, respectively, have a degenerate mass spectrum given by $M_{n}^2=4\pi^2\Lambda^2 n(n+1)$ for $n=1,2,...$.

\section{Conclusion}
 We have shown that the large black hole branch of the non-extremal rotating black 3-brane background solution (\ref{metric}) has pure Yang-Mills-like equation of state: the pressure $p$ vanishes at critical temperature $T_{c}=T_{min}=\sqrt{2}\Lambda$, see Fig.~\ref{fEoS}; the trace anomaly $\epsilon-3p$ have a maxima around $T_{c}$ and vanishes at very high temperature, see Fig.~\ref{facsym}; and the speed of sound $c_{s}^2$ approaches its conformal limit $1/3$ from below. In order to compare our results with pure Yang-Mills theory on the lattice and improved holographic QCD see Fig.5-9 in \cite{Gursoy:2010fj}.

%We have also shown that the small black hole branch of rotating black 3-branes is dynamically stable, and doesn't suffer from Gregory and Laflamme (GL) instablity unlike small Schwarzschild black hole with spherical horizon in $AdS_{5}\times S^5$. 

%And, we have conjectured that hadronization in $\mathcal{N}=4$ cSYM plasma is a holographic dual to Hawking radiation from the large black hole branch of rotating black 3-branes at freez-out temperature $T_{f}=T_{min}=\sqrt{2}\Lambda$. The holographic hadronization proposed in this paper can also be applied to other confining holographic models of QCD, such as hard/soft-wall AdS/QCD, improved holographic QCD, Sakai-Sugimoto model, etc.     

Note that we have investigated the hydrodynamic transport coefficients, and hard probe parameters of the strongly coupled $\mathcal{N}=4$ cSYM plasma in \cite{Mamo:2016oli}.

\section{Acknowledgements}

The author thanks Ho-Ung Yee for stimulating discussions and helpful comments on the
draft, Pablo Morales, Andrey Sadofyev, and Yi Yin for discussions.

 \vfil

\end{document}